\documentclass[12pt]{iopart}
\usepackage{graphicx}

\begin{document}

\title{Density anomaly in a  competing interactions lattice gas model}

\author{Alan B. de Oliveira }

\address{Instituto de F\'{\i}sica, Universidade Federal do Rio Grande do
Sul, Caixa Postal 15051, 91501-970, Porto Alegre, RS, Brazil}

\author{Marcia C. Barbosa\footnote[3]{To
whom correspondence should be addressed (barbosa@if.ufrgs.br)}}

\address{Instituto de Física, Universidade Federal do Rio Grande do
Sul, Caixa Postal 15051, 91501-970, Porto Alegre, RS, Brazil}

\begin{abstract}
We study  a very simple model of a short-range attraction and an
outer shell repulsion as a test system for 
demixing phase transition and density anomaly. The
phase-diagram is obtained by applying mean field analysis and Monte Carlo
simulations to  a two
dimensional  lattice gas
with nearest-neighbors attraction and next-nearest-neighbors repulsion (the 
outer shell). Two 
liquid phases and density anomaly are found. 
 The coexistence line between these two liquid phases 
meets a critical line between the fluid and 
the low density liquid  at a tricritical point. The  line 
of maximum density emerges in the vicinity of the tricritical point, close 
to the demixing transition.
\end{abstract}

\maketitle

%%%%%%%%%%%%%%%%%%%%%%%%%%%%%%
%%%%%%%%%%%%%%%%%%%%%%%%%%%%%%
\section{Introduction}
%%%%%%%%%%%%%%%%%%%%%%%%%%%%%
%%%%%%%%%%%%%%%%%%%%%%%%%%%%%%

Water is anomalous substance in many respects.

Most liquids contract upon cooling. This is not the case of water, a liquid
where the specific volume at ambient pressure starts to increase 
when cooled below $T=4 ^oC$ \cite{Wa64}. Besides, in a certain
range of pressures, also exhibits an anomalous increase of compressibility 
and specific heat upon cooling \cite{Pr87}-\cite{Ha84}. It is 
less well known that water diffuses faster under pressure at
very high densities and at
very low temperatures \cite{Ne00}. Besides,  the viscosity of water
decreases upon increasing pressure  \cite{De96}-\cite{Yo91}.

It was proposed a few years ago
that these anomalies are related to a second critical
point between two liquid phases, a low density liquid
(LDL) and a high density liquid (HDL) \cite{Po92}. This critical point 
was discovered
by computer simulations. This work   suggests that this
critical point is located at the 
supercooled region beyond  the line of
homogeneous nucleation and thus cannot be experimentally measured. 
Even if this limitation,  this hypothesis has been supported by indirect experimental 
results \cite{Mi98}\cite{angell}.

In spite of the limit of \( 235K \)
below which water cannot be found in the liquid phase without crystallization,
two amorphous phases were observed at much lower temperatures \cite{Mi84}.
There is evidence, although yet under test, that these two amorphous phases
are related to two liquid phases in  fluid water \cite{Sm99}\cite{Mi02}.

Water is not an isolated case. There are also other examples of  tetrahedrally
bonded molecular liquids such as phosphorus \cite{Ka00}\cite{Mo03}
and amorphous  silica \cite{La00} that also are good candidates for having
two liquid phases. Moreover, other materials such as liquid metals
\cite{Cu81} and graphite \cite{To97} also exhibit thermodynamic anomalies.
Unfortunately a coherent and general interpretation of 
 the 
low density liquid and high density liquid phases is still missing.

What type of potential would be appropriated for 
describing the tetrahedrally bonded 
molecular liquids? Directional interactions are  certainly an 
important ingredient in obtaining a quantitative  predictions
for network-forming liquids like water. However, the models
that are obtained  from that approach are too complicated, being
impossible to go beyond mean field analysis.
Isotropic models became the simplest framework to understand the physics of 
the liquid-liquid phase transition and liquid state anomalies. 

Recently it has been shown that the presence of two liquid phases
can be associated with a potential with an attractive part and
two characteristic short-range 
repulsive distances. The smallest of these two distances is 
associated with the hard core of the molecule, while the largest
one with the soft core \cite{Fr01}.
Acknowledging that core softed (CS) potentials may engender a demixing
transition between two liquids of different densities, a number of 
CS potentials were proposed to model the anisotropic systems described above.
The first suggestion was made many years ago by Stell and coworkers in order to explain the isostructural solid-solid
transition ending in a critical point\cite{He70}-\cite{Ho73}.
Debenedetti et al. using general thermodynamic arguments, confirmed 
that a CS can lead to a coefficient of thermal expansion negative and
consequently to density anomaly \cite{De91}. This together
with the increase of the thermal compressibility has
been used as indications of the presence of two liquid
phases \cite{St98}\cite{St00} which may be hidden beyond the homogeneous
nucleation.  The difficulty  with these approaches is 
that continuous potentials usually lead to crystallization 
at the region where the critical point would be expected. The 
analysis of the presence of both the two liquid phases and the
critical point becomes indirect.

Nevertheless the apparent success of the CS potentials, it is 
not clear that the presence of two liquid phases and anomalies 
are associated only with them  \cite{De91} or if they 
result from the presence of 
 two competing scales like the ones present in
the CS potentials. 
In this work we analyze another type of 
model system  where two competing scales are also
present. We study a potential  with a hard core, a short-range
 attractive part and
a repulsive shell. While the attraction
accounts both for the van der Waals and hydrogen bonding 
interactions, the outer shell repulsion is related to the 
 interstitial molecule that break the tetrahedral structure 
like the interstitial oxygens in water. In order to circumvent
crystallization, our model system is a two dimensional lattice 
gas with first-neighbors attraction and second neighbors repulsion.
The system is in contact with a reservoir of particles.

We show that this very simple
system exhibits  both density anomaly and two liquid phases.
However, instead of 
having a critical point ending the coexistence line between
the two liquid phases as one usually would expect, it has a tricritical point.
The connection between the presence of criticality and the density
anomaly shown. 

The reminder of the paper goes as follows. In sec.(\ref{2}) the model is presented, the mean field analysis is shown on
sec.(\ref{3}), results from simulations are discussed in
sec.(\ref{4}) and our findings are summarized in sec.(\ref{5}).

%%%%%%%%%%%%%%%%%%%
%%%%%%%%%%%%%%%%%%
\section{The model}
\label{2}
%%%%%%%%%%%%%%%%%%%
%%%%%%%%%%%%%%%%%

\begin{figure}
\begin{center}
\begin{minipage}[c]{0.80\columnwidth}
\begin{center}
\includegraphics[scale=0.5]{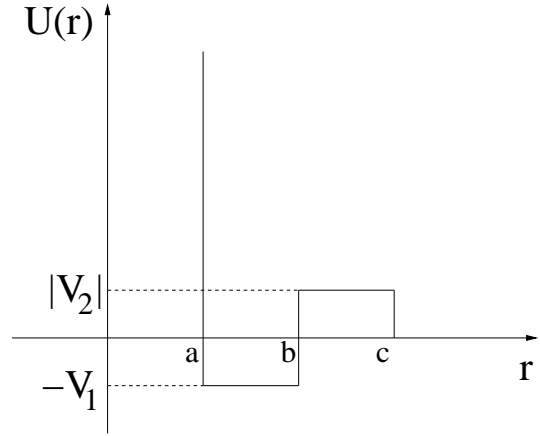}
\end{center}
\caption{Intermolecular interaction \label{potential}}
\end{minipage}
\end{center}
\end{figure}

\begin{figure}
\begin{center}
\begin{minipage}[c]{0.60\columnwidth}
\begin{center}
\includegraphics[bb=0bp 0bp 316bp 315bp,clip,scale=0.4]{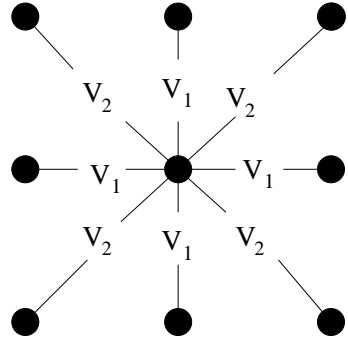}
\end{center}
\caption{The neighbor's interactions \label{vizinhos}}
\end{minipage}
\end{center}
\end{figure}

Our system is represented by a square lattice with $N$ sites. Associated
to each site there is an occupational variable, $\sigma_i$. If the  site 
is occupied by molecule,  $\sigma_{i}=1$, otherwise
$\sigma_{i}=0$.
Each site interacts with its nearest neighbors with attractive interactions
and with its next-nearest neighbors with repulsive interactions ( see Fig.\ref{potential}). Therefore
the Hamiltonian  of this system is given by
%%%%%%%%%%%%%%%%%%%%%%%%%%%%%%%%%%%%%%%%%%%%%%%%%%%%%%%%%%%%%%%%%%%%%
\begin{equation}
{\cal H}=-V_{1}\sum_{<ij>}\sigma_{i}\sigma_{j}-V_{2}\sum_{<<ik>>}\sigma_{i}\sigma_
{k}-\mu\sum_{i}\sigma_{i},
\label{hamil}
\end{equation}
%%%%%%%%%%%%%%%%%%%%%%%%%%%%%%%%%%%%%%%%%%%%%%%%%%%%%%%%%%%%%%%%%%%%
\noindent where the first sum is over the four nearest-neighbors,
with energy gain $V_{1}>0$, and the second sum, over the four next-nearest-neighbors,
has energy interaction $V_{2}<0$ (see Fig.\ref{vizinhos}).  The last 
term in eq.(\ref{hamil}) refers
to chemical potential contribution $\mu$ over all molecules. Here we will
consider periodic boundary conditions. 

%%%%%%%%%%%%%%%%%%%%%%%%%%%%%%%%%%%%%%%%%%%%%%%%%%%%%%%%%%
%%%%%%%%%%%%%%%%%%%%%%%%%%%%%%%%%%%%%%%%%%%%%%%%%%%%%%%%%%
%%%%%%%%%%%%%%%%%%%%%%%%%%%%%%%%%%%%%%%%%%%%%%%%%%%%%%%%%%%%%%%%%
\begin{figure}
\begin{center}
\begin{minipage}[c]{0.60\columnwidth}
\begin{center}
\includegraphics[scale=0.34]{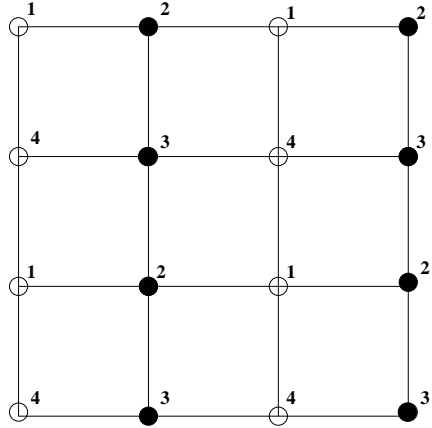}
\end{center}
\caption{Sub-lattices \label{sub-redes}}
\end{minipage}
\end{center}
\end{figure}

%%%%%%%%%%%%%%%%%%%%%%%%%%%%%%%%%%%%%%%%%%%%%%%%%%%%
In order to help visualization of the possible  phases,
the lattice is divided  into four sub-lattices: $1,2,3,4$ (appropriated
for the description of an arbitrary superstructure with twice the
lattice spacing of the original lattice). The corners of a simple
square are labeled counter-clockwise to indicate the sub-lattices
(see example in Fig.\ref{sub-redes}.)

The sub-lattice density is given by
%%%%%%%%%%%%%%%%%%%%%%%%%%%%%%%%%%%%%%%%%%%%%%%%%%%%%%%%%%
\begin{equation}
\rho_{\beta}=\frac{4}{N}\sum_{j\in\beta}\sigma_{j},\,\,\,\,\,
\beta=1,\ldots,4.
\label{eq:densidade}
\end{equation}
%%%%%%%%%%%%%%%%%%%%%%%%%%%%%%%%%%%%%%%%%%%%%%%%%%%%%%%%%%

%%%%%%%%%%%%%%%%%%%%%%%%%%%%%%%%%%%%%%%%%%%%%%%%%%%%%%%%%%%
%%%%%%%%%%%%%%%%%%%%%%%%%%%%%%%%%%%%%%%%%%%%%%%%%%%%%%%%%%%

The ground state is defined by the lower  grand potential free energy at
$T=0$, and therefore the minimum value of Hamiltonian as given
by eq.(\ref{hamil}). Due the symmetry of the system, in the  case
$V_{1}>0$ and $V_{2}<0,$ there are  two possibilities illustrated below.

\subsubsection{If $V_{1}<-2V_{2}$}

\medskip

For $\mu>-3V_{1}-4V_{2}$ the system  stands in the the  dense liquid phase (DL),
where all sites are  occupied. As  the chemical potential is decreased,at $\mu=-3V_{1}-4V_{2}$ 
there is  a phase transition from the dense liquid phase  to
the  structured dilute liquid phase (SDL), illustrated
in Fig.\ref{sub-redes}. Decreasing the chemical potential 
even further, the structured
dilute liquid persists until $\mu=-V_{1},$ where  there is  a phase
transition between the structured liquid and the  gas phase.
for $\mu<-V_{1}$, the system stays empty.

\subsubsection{If $V_{1}>-2V_{2}$}

\medskip

For $\mu>-2V_{1}-2V_{2}$ the system is in the
 dense liquid and for  $\mu<-2V_{1}-2V_{2}$
the stable one is the gas phase. In this case
only one phase transition  
at $\mu=-2V_{1}-2V_{2}$ is present.

%%%%%%%%%%%%%%%%%%%%%%%%%%%%%%%%%%%
%%%%%%%%%%%%%%%%%%%%%%%%%%%%%%%%%
\section{Mean-Field Approximation}
\label{3}
%%%%%%%%%%%%%%%%%%%%%%%%%%%%%%%%%%%
%%%%%%%%%%%%%%%%%%%%%%%%%%%%%%%%%%%

For the mean field analysis, we will follow a sub-lattice strategy
\cite{Bi80} that it is able to capture the phase transition that 
other mean field schemes miss.
The chemical potential can be rewritten as the sum of four potentials, one
for  each sub-lattice:
%%%%%%%%%%%%%%%%%%%%%%%%%%%%%%%%%%%%%%%%%%%%%%%%%%%%%%%%%%%%%%%%%%%%%
\begin{equation}
\mu=\frac{1}{4}\sum_{\alpha=1}^{4}\mu_{\alpha}.
\label{eq:pot}
\end{equation}
%%%%%%%%%%%%%%%%%%%%%%%%%%%%%%%%%%%%%%%%%%%%%%%%%%%%%%%%%%%%%%%%%%%%%

The use of independent chemical potentials may be necessary in a more
general problem, where each sub-lattice is composed by a different
type of molecule. For our model we have:
%%%%%%%%%%%%%%%%%%%%%%%%%%%%%%%%%%%%%%%%%%%%%%%%%%%%%%%%%%%%%%%%%%
\begin{equation}
\mu_{1}=\mu_{2}=\mu_{3}=\mu_{4},
\label{eq:pot2}
\end{equation}
%%%%%%%%%%%%%%%%%%%%%%%%%%%%%%%%%%%%%%%%%%%%%%%%%%%%%%%%%%%%%%%%
which leads, by eq.(\ref{eq:pot}), to\[\mu=\mu_{\alpha},\,\,\,\,\alpha=1,2,3,4.\]

Now, rewriting the Hamiltonian , eq.(\ref{hamil}), in term of
the sub-lattices,  we get
%%%%%%%%%%%%%%%%%%%%%%%%%%%%%%%%%%%%%%%%%%%%%%%%%%%%%%%%%%%%%%%%
\begin{equation}
{\cal H}=-\sum_{\alpha=1}^{4}\sum_{i\in\alpha}H_{\alpha}^{eff}\left(\left\{ \sigma_{j}\right\} \right)\sigma_{i},
\label{hamilmedio}
\end{equation}
%%%%%%%%%%%%%%%%%%%%%%%%%%%%%%%%%%%%%%%%%%%%%%%%%%%%%%%%%%%%%%%%
where
%%%%%%%%%%%%%%%%%%%%%%%%%%%%%%%%%%%%%%%%%%%%%%%%%%%%%%%%%%%%%%%%
\[
H_{\alpha}^{eff}\left(\left\{ \sigma_{j}\right\} \right)=\mu_{\alpha}+\sum_{\beta=1}^{4}\sum_{i\in\beta}J_{ij}\sigma_{j},\,\,\,\, i\in\alpha\]
%%%%%%%%%%%%%%%%%%%%%%%%%%%%%%%%%%%%%%%%%%%%%%%%%%%%%%%%%%%%%%%%
and where the relation between $J_{ij},V_{1}$ and $V_{2}$ is given by:
%%%%%%%%%%%%%%%%%%%%%%%%%%%%%%%%%%%%%%%%%%%%%%%%%%%%%%%%%%%%%%%%
\[
J_{12}=J_{21}=J_{23}=J_{32}=J_{34}=J_{43}=J_{14}=J_{41}=V_{1}\]
\[
J_{13}=J_{31}=J_{24}=J_{42}=V_{2}.\]
%%%%%%%%%%%%%%%%%%%%%%%%%%%%%%%%%%%%%%%%%%%%%%%%%%%%%%%%%%%%%%%%

The  mean-field approximation here  consists in replacing  $H_{\alpha}^{eff}$
by its  mean value, given by
%%%%%%%%%%%%%%%%%%%%%%%%%%%%%%%%%%%%%%%%%%%%%%%%%%%%%%%%%%%%%%%%
\begin{equation}
\left\langle H_{\alpha}^{eff}\right\rangle =\mu_{\alpha}+\sum_{\beta=1}^{4}\sum_{i\in\beta}J_{ij}\left\langle \sigma_{j}\right\rangle ,\,\,\,\, i\in\alpha.
\label{eq:effmedio}
\end{equation}
%%%%%%%%%%%%%%%%%%%%%%%%%%%%%%%%%%%%%%%%%%%%%%%%%%%%%%%%%%%%%%%%
Under this approximation,  the density of the sub-lattice $\beta$ is given by
$\rho_{\beta}=\left\langle \sigma_{j}\right\rangle $ and the interaction
parameter can be writen as
%%%%%%%%%%%%%%%%%%%%%%%%%%%%%%%%%%%%%%%%%%%%%%%%%%%%%%%%%%%%%%%%
\[
\epsilon_{\alpha\beta}=\sum_{i(\neq j)}J_{ij},\,\,\, i\in\alpha,\,\,\, j\in\beta,\]
%%%%%%%%%%%%%%%%%%%%%%%%%%%%%%%%%%%%%%%%%%%%%%%%%%%%%%%%%%%%%%%%
what leads to 
%%%%%%%%%%%%%%%%%%%%%%%%%%%%%%%%%%%%%%%%%%%%%%%%%%%%%%%%%%%%%%%%
\begin{equation}
\left\langle H_{\alpha}^{eff}\right\rangle =\mu_{\alpha}+\sum_{\beta=1}^{\nu}\epsilon_{\alpha\beta}\rho_{\beta},\,\,\,\, i\in\alpha,\,\, j\in\beta.
\label{eq:effmedio2}
\end{equation}
%%%%%%%%%%%%%%%%%%%%%%%%%%%%%%%%%%%%%%%%%%%%%%%%%%%%%%%%%%%%%%%%

Substituting eq.(\ref{eq:effmedio2}) in eq.(\ref{hamilmedio})
we get 
%%%%%%%%%%%%%%%%%%%%%%%%%%%%%%%%%%%%%%%%%%%%%%%%%%%%%%%%%%%%%%%%
\begin{equation}
{\cal H}^{MF}=-\sum_{\alpha=1}^{4}\sum_{i\in\alpha}\left(\sum_{\beta=1}^{4}\epsilon_{\alpha\beta}\rho_{\beta}+\mu_{\alpha}\right)\sigma_{i}+\frac{1}{2}\sum_{\alpha=1}^{4}\frac{N}{4}\sum_{\beta=1}^{4}\epsilon_{\alpha\beta}\rho_{\beta}\rho_{\alpha},
\label{hamilmedio2}
\end{equation}
%%%%%%%%%%%%%%%%%%%%%%%%%%%%%%%%%%%%%%%%%%%%%%%%%%%%%%%%%%%%%%%%
where the last term corrects for overcounting. Eq.(\ref{hamilmedio2}) is 
the Hamiltonian in the mean-field approximation.
With this, we calculate the mean-field  grand potential per site,
that is
%%%%%%%%%%%%%%%%%%%%%%%%%%%%%%%%%%%%%%%%%%%%%%%%%%%%%%%%%%%%%%%%
\begin{eqnarray}
\phi^{MF} & = & -k_{B}T\ln2\nonumber \\
 &  & -\frac{k_{B}T}{4}\sum_{\alpha=1}^{4}\ln\cosh\left[\frac{1}{2k_{B}T}\left(\sum_{\beta=1}^{4}\epsilon_{\alpha\beta}\rho_{\beta}+\mu_{\alpha}\right)\right]\nonumber \\
 &  & -\frac{1}{8}\sum_{\alpha=1}^{4}\left(\sum_{\beta=1}^{4}\epsilon_{\alpha\beta}\rho_{\beta}+\mu_{\alpha}\right)+\frac{1}{8}\sum_{\alpha=1}^{4}\sum_{\beta=1}^{4}\epsilon_{\alpha\beta}\rho_{\alpha}\rho_{\beta},\label{eq:potencial}\end{eqnarray}
%%%%%%%%%%%%%%%%%%%%%%%%%%%%%%%%%%%%%%%%%%%%%%%%%%%%%%%%%%%%%%%%
where $k_{B}$ is the  Boltzmann factor  and $T$ is the  temperature. 

The density can be derived from this grand potential as:
%%%%%%%%%%%%%%%%%%%%%%%%%%%%%%%%%%%%%%%%%%%%%%%%%%%%%%%%%%%%%%%%
\[
\rho_{\alpha}=-4\left(\frac{\partial\phi^{MF}}{\partial\mu_{\alpha}}\right)_{T,\mu_{\alpha\neq\beta}},\,\,\,\,\alpha=1,2,3,4,\]
%%%%%%%%%%%%%%%%%%%%%%%%%%%%%%%%%%%%%%%%%%%%%%%%%%%%%%%%%%%%%%%%
what leads to 
%%%%%%%%%%%%%%%%%%%%%%%%%%%%%%%%%%%%%%%%%%%%%%%%%%%%%%%%%%%%%%%%
\begin{equation}
\rho_{\alpha}=\frac{1}{2}+\frac{1}{2}\tanh\left[\frac{1}{2k_{B}T}\left(\sum_{\beta=1}^{4}\epsilon_{\alpha\beta}\rho_{\beta}+\mu_{\alpha}\right)\right],\,\,\,\,\alpha=1,2,3,4,
\label{eq:densidades}
\end{equation}
%%%%%%%%%%%%%%%%%%%%%%%%%%%%%%%%%%%%%%%%%%%%%%%%%%%%%%%%%%%%%
the mean density of a sub-lattice $\alpha.$

\begin{figure}
\begin{center}
\includegraphics[ bb=136bp 42bp 556bp 352bp,  clip, scale=0.7]{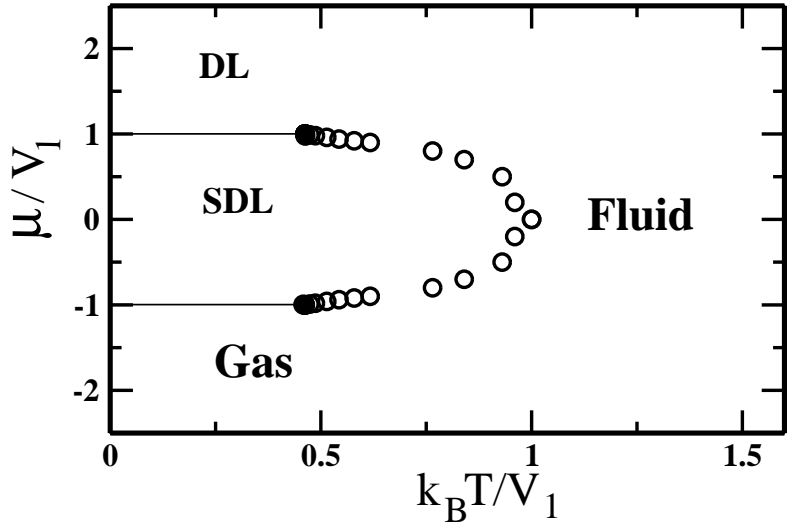}
\end{center}
\caption{Mean field phase diagram \label{diagramacm}: The open  circles represent  the second order phase transition, the solid  lines are  the two  first-order phase transitions and  
the two filled circles are the two tricritical points.}
\end{figure}

Solving Eqs.~(\ref{eq:densidades}) for fixed values 
of temperature and chemical potential it is possible to obtain
the density of each sub-lattice and consequently not only  the
overall density of the system but also  specific phase. For instance, 
 if for a given temperature and chemical potential, the
sub-lattice densities would be   $\rho_{1}=\rho_{4}=0$ 
and $\rho_{2}=\rho_{3}=1$,
the system would be  in structured dilute liquid phase. 
Since we know that for $T=0$ two liquid phases exist only
if $V_1<-2V_2$, we explore this region of the parameter 
space. For simplicity, we set  $V_{1}=1$ and $V_{2}=-1$ (other parameter
choices will not affect qualitatively the result) and 
we construct the mean-field phase
diagram shown in Fig. \ref{diagramacm}.

%%%%%%%%%%%%%%%%%%%%%%%%%%%%%%%%%%%%%%%%%%%%%%%%%%%%%%%%%%%%
%%%%%%%%%%%%%%%%%%%%%%%%%%%%%%%%%%%%%%%%%%%%%%%%%%%%%%%%%%%%

At high temperatures, each sub-lattice is half full in an disorganized 
way. This is the fluid phase. As the temperature is lower at a fixed
potential $=\mu^*=\mu/V_1>1$, all sub-lattices become filled, no 
phase transition
is observed. For
very low chemical potentials,  $\mu^*<-1$, as the temperature is decreased, 
the system goes from the fluid to the gas phase continuously with
no phase transition. For intermediate
chemical potentials, $-1>\mu^*>1$, there is a continuous phase transition
between the fluid phase and the structured dilute liquid phase. The 
coexistence line between the gas phase and the structured diluted 
liquid meets the continuous phase transition at a tricritical
point. Another similar point is observed at the contact 
between the coexistence line between the structured diluted
liquid and the dense liquid and the continuous line. The
density is monotonic, therefore no density anomaly is observed.

%%%%%%%%%%%%%%%%%%%%%%%%%%%%%%%%%
%%%%%%%%%%%%%%%%%%%%%%%%%%%%%%%%%
\section{Monte Carlo Simulations}
\label{4}
%%%%%%%%%%%%%%%%%%%%%%%%%%%%%%%%%
%%%%%%%%%%%%%%%%%%%%%%%%%%%%%%%%%

The rather simple mean field approach introduced  in the previous
session is unable to account for the density anomalies. For
investigating the possibility of a density anomaly in
our  potential,  Monte Carlo simulations in
grand canonical ensemble were performed. 
The Metropolis algorithm was used to study square
$L\times L$ lattice and $|V_2|/V_1=1$.  Different
system sizes $L=10,20,30,40,50$ were investigated. The typical equilibration
time was $1 500 000 $ Monte Carlo time steps for each lattice site.

The nonzero temperature phase-diagram was obtained as follows. For
a fixed temperature and chemical potential the density of each
sub-lattice and the specific heat was computed by averaging over
5000 measures. Between consecutive measures, $\tau$ Monte Carlo
steps were performed to decorrelated the system. The 
 \emph{correlation time $\tau$} was  calculated using
the density correlation function \cite{key-1}:
%%%%%%%%%%%%%%%%%%%%%%%%%%%%%%%%%%%%%%%%%%%%%%%%%%%%%%%%%%%%%%%%%%%%%%%%
\begin{eqnarray}
\chi(t) & = & \frac{1}{t_{max}-t}\sum_{\tilde{t}=0}^{t_{max}-t}\rho(\tilde{t})\rho(\tilde{t}+t)-\nonumber \\
 &  & \left[\frac{1}{t_{max}-t}\sum_{\tilde{t}=0}^{t_{max}-t}\rho(\tilde{t})\right]\left[\frac{1}{t_{max}-t}\sum_{\tilde{t}=0}^{t_{max}-t}\rho(\tilde{t}+t)\right].
\label{eq:correlacao}
\end{eqnarray}
%%%%%%%%%%%%%%%%%%%%%%%%%%%%%%%%%%%%%%%%%%%%%%%%%%%%%%%%%%%%%%%%%%%%%%%%
%%%%%%%%%%%%%%%%%%%%%%%%%%%%%%%%%%%%%%%%%%%%%%%%%%%%%%%
\begin{figure}
\begin{center}
\includegraphics[bb=187bp 249bp 724bp 610bp,clip,scale=0.5]{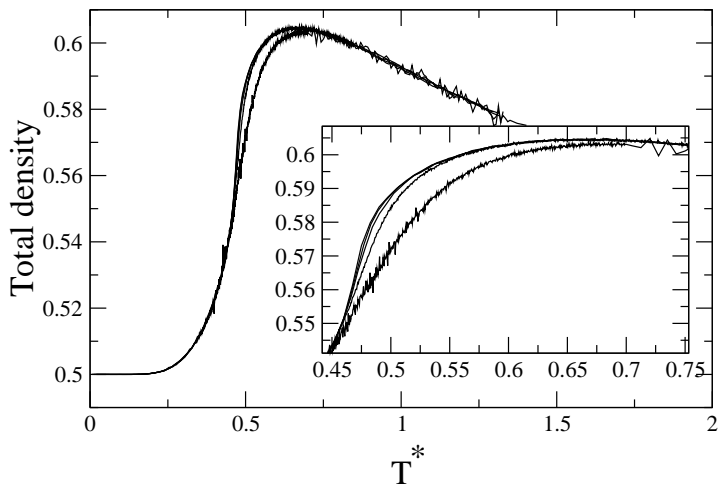}
\end{center}
\caption{Total density \label{d-total}:for the 50, 40, 30, 20 and 10 lattices, from
top to bottom. }
\end{figure}
%%%%%%%%%%%%%%%%%%%%%%%%%%%%%%%%%%%%%%%%%%%%%%%%%%%%%%%

\begin{figure}
\begin{center}
\begin{minipage}[c]{0.45\columnwidth}
\begin{center}
\includegraphics[ clip, scale=0.425]{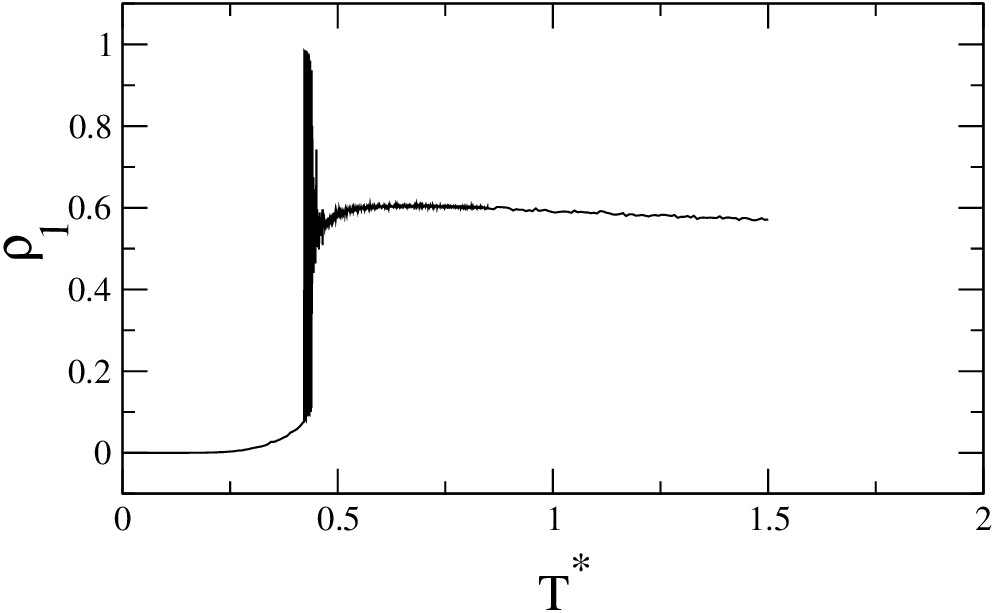}
\end{center}
\caption{Sub-lattice 1 \label{sub1}}
\end{minipage}
\begin{minipage}[c]{0.45\columnwidth}
\begin{center}
\includegraphics[clip,scale=0.425]{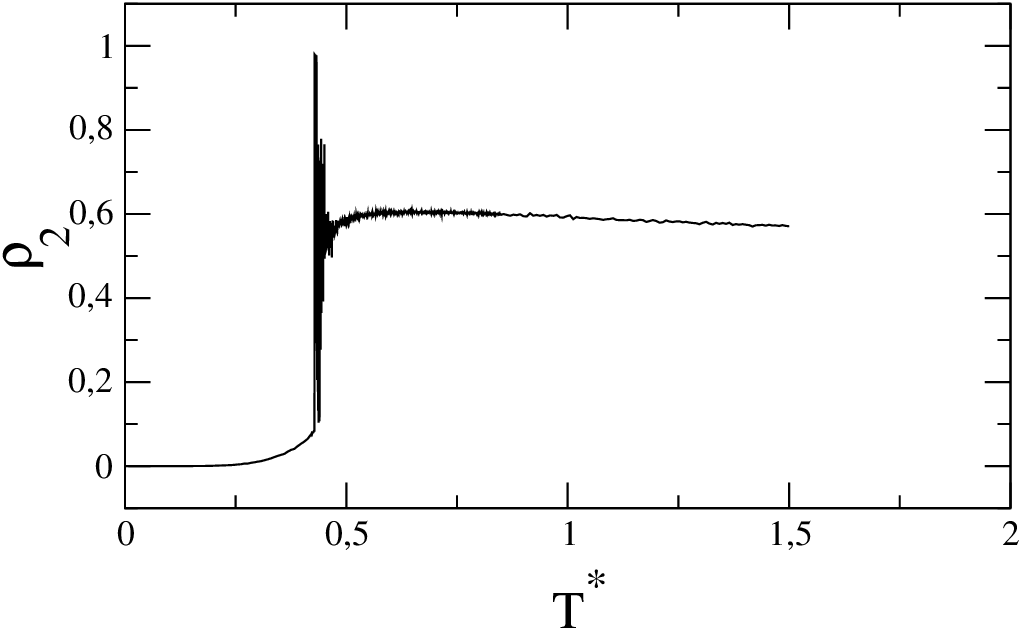}
\end{center}
\caption{Sub-lattice 2 \label{sub2}}
\end{minipage}
\end{center}
\end{figure}

\begin{figure}
\begin{center}
\begin{minipage}[c]{0.45\columnwidth}%
\begin{center}
\includegraphics[ clip, scale=0.425]{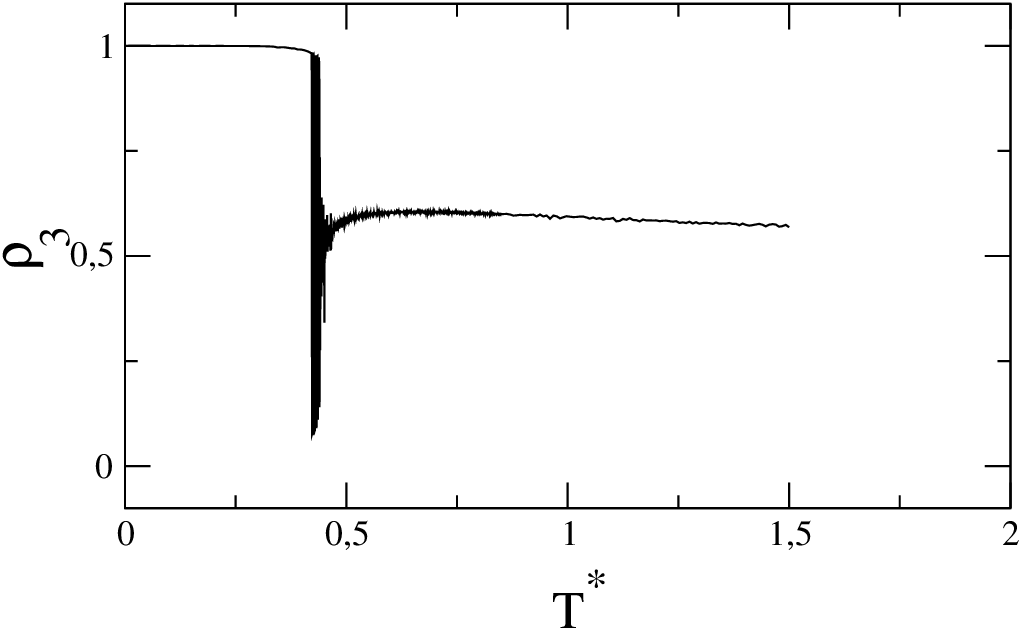}
\end{center}
\caption{Sub-lattice 3 \label{sub3}}
\end{minipage}
\begin{minipage}[c]{0.45\columnwidth}
\begin{center}
\includegraphics[clip,scale=0.425]{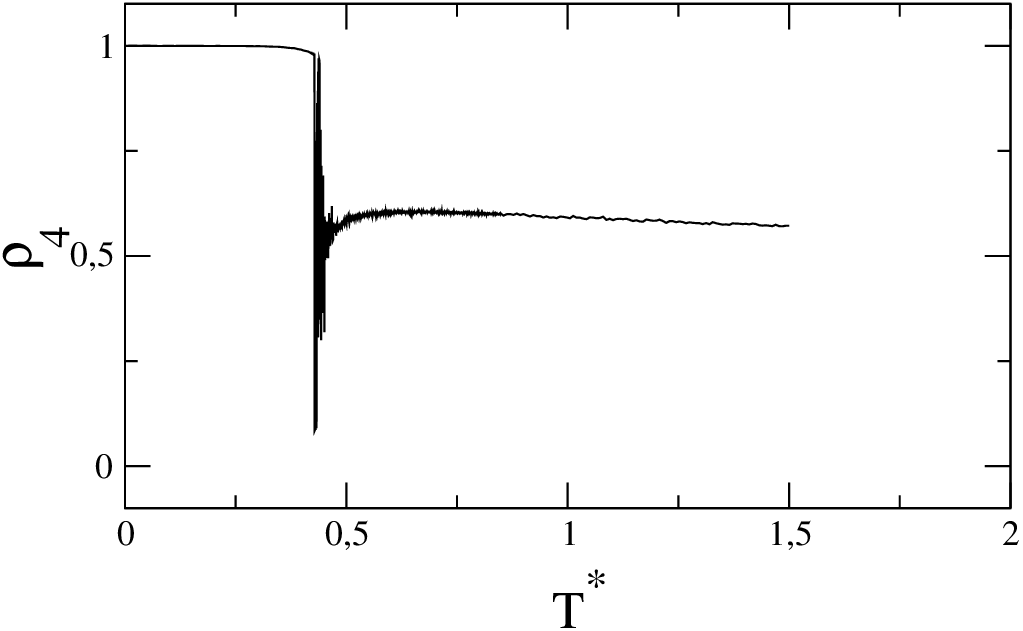}
\end{center}
\caption{Sub-lattice 4 \label{sub4}}
\end{minipage}
\end{center}
\end{figure}

\begin{figure}
\begin{center}
\begin{center}
\includegraphics[ clip, scale=0.55]{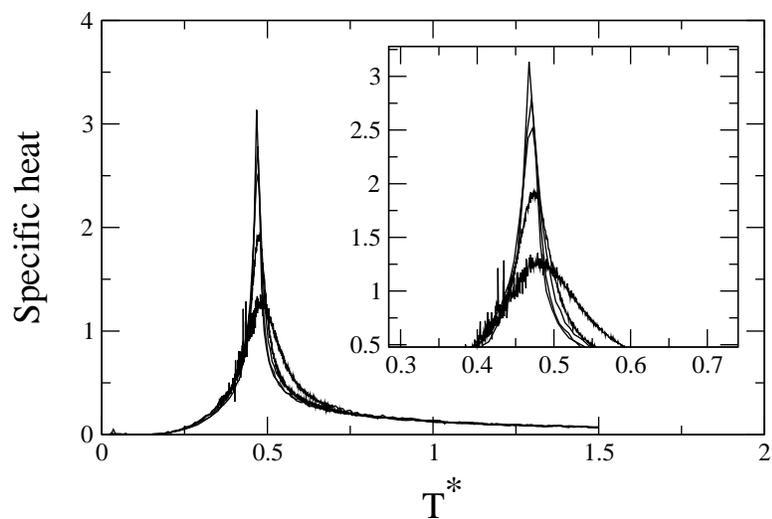}
\end{center}
\caption{Specific heat for  $\mu^*=0.50$ \label{calesp-todos}and  $L=50, 40,
30, 20,10$ from top to bottom.}
\end{center}
\end{figure}

Fig. \ref{d-total} shows the total density for  the lattice sizes  $L=50,
40, 30, 20$ and $10$, from top to bottom (see also inset) and for 
fixed chemical potential $\mu^*=0.5$. 
According to this graph, one could conclude that 
no phase transition happens as the temperature is 
decreased at a fixed chemical potential $\mu^*=0.5$. However, analyzing 
the sub-lattice's
densities illustrated in 
Figs. \ref{sub1} - \ref{sub4} ( here we are illustrating only the
result obtained for the $L=20$), it's clear from the strong density 
fluctuations that appear  at  $T^{*}=T/V_1=0.45$ a 
phase transition occurs. The peak  in the specific heat shown in Fig. \ref{calesp-todos} confirms the 
existence of this transition and helps to determined 
 the exact location of the 
critical point  $T^{*}_c$.

%%%%%%%%%%%%%%%%%%%%%%%%%%%%%%%%%%%%%%%%%%%%%%%%%%%%%%%%
%%%%%%%%%%%%%%%%%%%%%%%%%%%%%%%%%%%%%%%%%%%%%%%%%%%%%%%%

%%%%%%%%%%%%%%%%%%%%%%%%%%%%%%%%%%%%%%%%%%%%%%%%%%%%%%%%%
%%%%%%%%%%%%%%%%%%%%%%%%%%%%%%%%%%%%%%%%%%%%%%%%%%%%%%%%%

%%%%%%%%%%%%%%%%%%%%%%%%%%%%%%%%%%%%%%%%%%%%%%%%%%%%%%%%%%
%%%%%%%%%%%%%%%%%%%%%%%%%%%%%%%%%%%%%%%%%%%%%%%%%%%%%%%%%%

\begin{figure}
\begin{center}
\includegraphics[clip,scale=0.5]{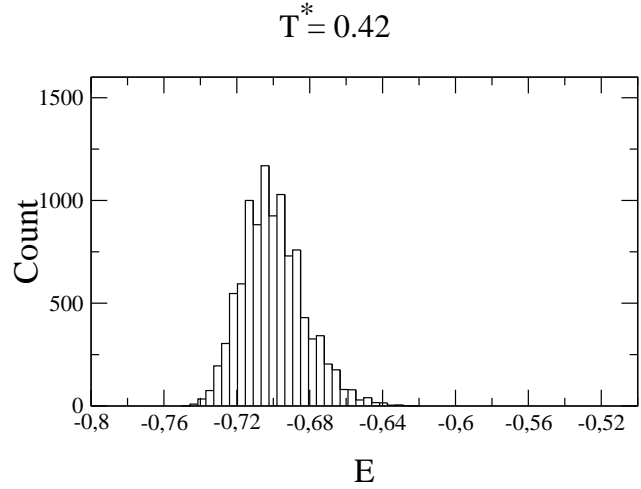}
\caption{Energy histogram for $T<T_c$,  $\mu^{*}=0.50$ and $L=20$
\label{histograma1}}
\end{center}
\end{figure}

\begin{figure}
\begin{center}
\includegraphics[clip, scale=0.5]{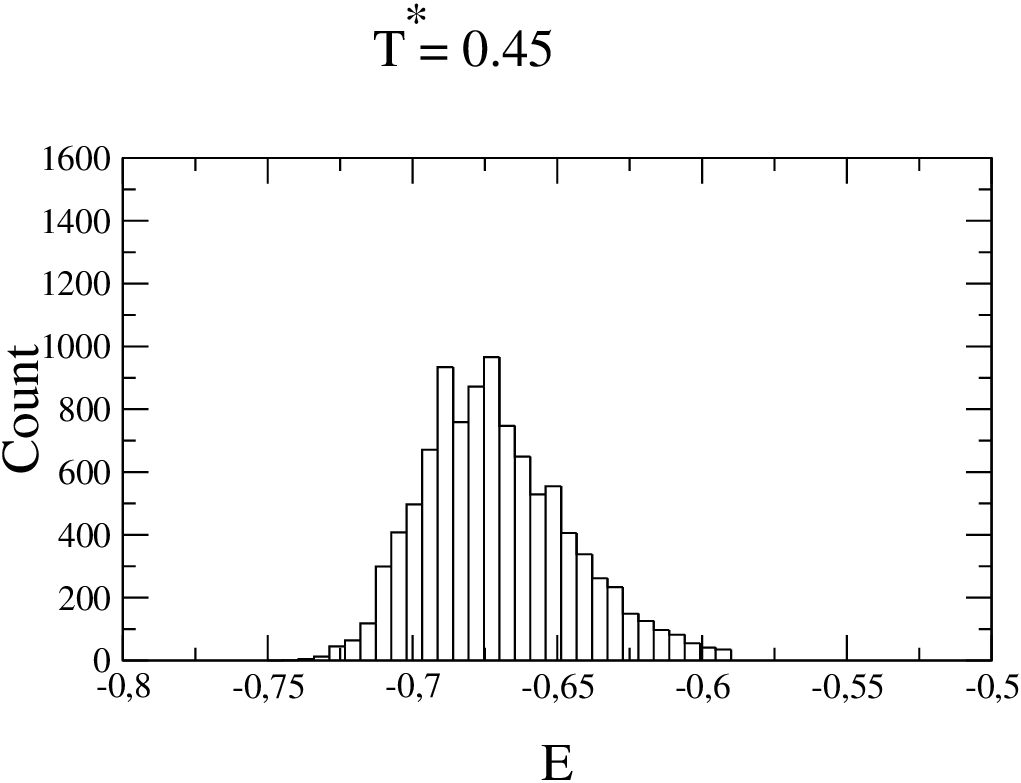}
\caption{Energy histogram for $T\approx T_c$,  $\mu^{*}=0.50$ and $L=20$\label{histograma2}}
\end{center}
\end{figure}

\begin{figure}
\begin{center}
\includegraphics[clip,scale=0.5]{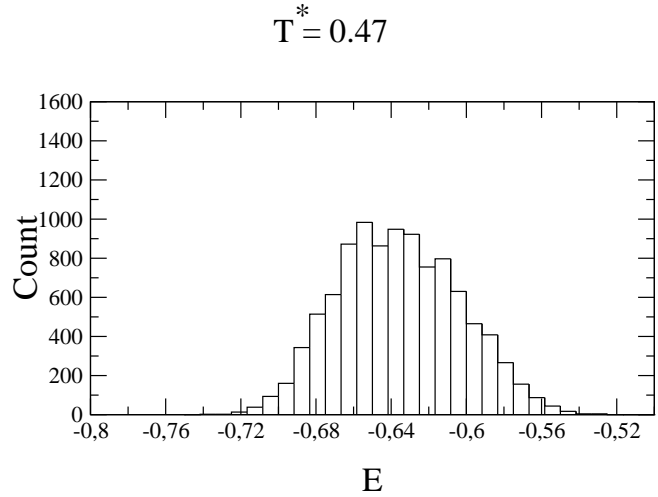}
\caption{Energy histogram for $T>T_c$,  $\mu^{*}=0.50$ and $L=20$\label{histograma3}}
\end{center}
\end{figure}

The energy histograms constructed for fixed temperatures
around the critical temperature $T_{c}^*$ are
 shown in Figs. \ref{histograma1}-\ref{histograma3}. For temperatures 
above and below $T_{c}^*$, there is only one
peak indicating that the transition is continuous ( two peaks would be a signature of first-order transition). 

 Analyzing again Figs. \ref{sub1}
- \ref{sub4} it can be seen  that for  $T^{*}>0.45$  the
sub-lattice densities   $\rho_{\beta}\approx 0.56$,  indicating 
that the system is in the  
 fluid phase.
For $T^{*}<0.45$,  $\rho_{1}=\rho_{2}=0$ and $\rho_{3}=\rho_{4}=1$,
that is a signature of the structured dilute liquid. Hence, it
is clear that around $T^*_c=0.45$ there is a 
continuous transition between the fluid and the SDL phases.
The error bars
associated  with  the location of the critical temperatures
for the lattice sizes  $L=10,20$ are respectively $\Delta T=4\times10^{-4}$,
 $\Delta T=7.5\times10^{-3}$ while for the lattice size  $L=30,40,50$ is
given by  $\Delta T=6.5\times10^{-3}$.

%%%%%%%%%%%%%%%%%%%%%%%%%%%%%%%%%%%%%%%%%%%%%%%%%%%%%%%%%%%%%%%%%
%%%%%%%%%%%%%%%%%%%%%%%%%%%%%%%%%%%%%%%%%%%%%%%%%%%%%%%%%%%%%%

%%%%%%%%%%%%%%%%%%%%%%%%%%%%%%%%%%%%%%%%%%%%%%%%%%%%%%%%
%%%%%%%%%%%%%%%%%%%%%%%%%%%%%%%%%%%%%%%%%%%%%%%%%%%%%%%%%%%%%%

%%%%%%%%%%%%%%%%%%%%%%%%%%%%%%%%%%%%%%%%%%%%%%%%%%%%%%%%
%%%%%%%%%%%%%%%%%%%%%%%%%%%%%%%%%%%%%%%%%%%%%%%%%%%%%%%%%%%%%%

 Criticality happens only in the thermodynamic limit. Therefore
in order to determined the actual critical temperature, it is necessary to
 extrapolate the results obtained for 
the finite system  to the limit of $L\rightarrow \infty$. 
The critical temperature for the finite system
can be obtained: from 
the maximum of the 
specific heat or from the minimum of the  fourth order Binder's cumulant 
 $V_{E}$ \cite{Key-2}.
This last method requires computing for each lattice size, the 
quantity:
%%%%%%%%%%%%%%%%%%%%%%%%%%%%%%%%%%%%%%%%%%%%%%%%%%%%%%%%%%%%%%%%%%
\begin{equation}
V_{E}=1-\frac{\left\langle E^{4}\right\rangle }{3\left\langle E^{2}\right\rangle ^{2}}\; 
\label{eq:cumbin}
\end{equation}
%%%%%%%%%%%%%%%%%%%%%%%%%%%%%%%%%%%%%%%%%%%%%%%%%%%%%%%%%%%%%%%%%%
where $E$ is the total energy of the system.

\begin{figure}
\begin{center}
\includegraphics[clip,scale=0.54]{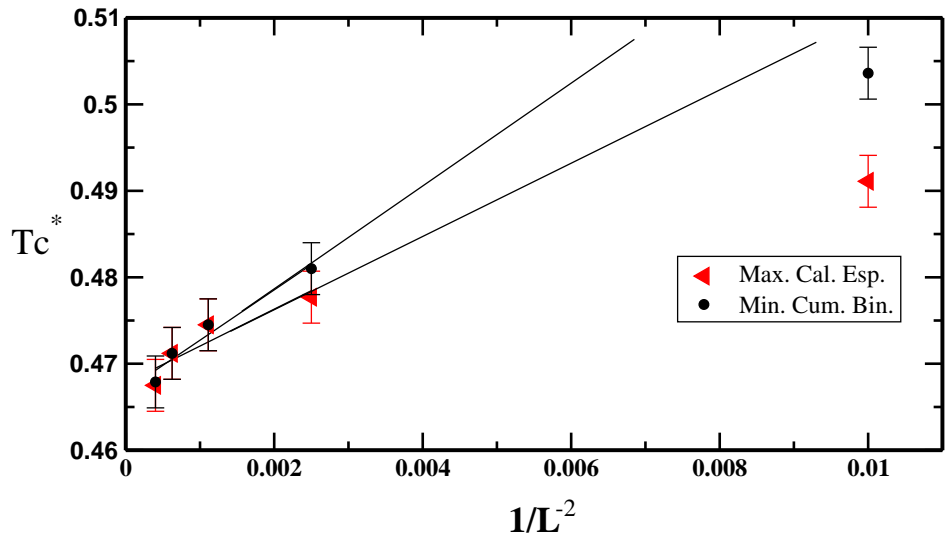}
\end{center}
\caption{ $T_c^* \times 1/L^{2}$ for $L=10,20,30,40,50$. The circles are the minimum of Binder's cumulant and the triangles are the maximum of specific heat. \label{extrapolating}}
\end{figure}

Fig. \ref{extrapolating} shows the critical temperature
of the finite system for different lattice sizes obtained from the 
two methods refereed above. The extrapolation to $1/L\rightarrow 0$
gives the critical temperature for the infinite system. The values
obtained from the peak of the specific heat and from
the minimum of the Binder's forth order cumulant are  respectively $0.467$
and $0.466$. The difference between the two results is 
smaller  than the error on the calculation of both numbers. The
lattice $L=10$ was excluded from the extrapolation 
 because it is so far from the infinite size limit that 
its inclusion would require carrying  nonlinear terms into
the extrapolation.

%%%%%%%%%%%%%%%%%%%%%%%%%%%%%%%%%%%%%%%%%%%%%%%%%%%%%%%%%%%%%%%%%%%
%%%%%%%%%%%%%%%%%%%%%%%%%%%%%%%%%%%%%%%%%%%%%%%%%%%%%%%%%%%%%%%%%%

\begin{figure}
\begin{center}
\includegraphics[clip,scale=0.6]{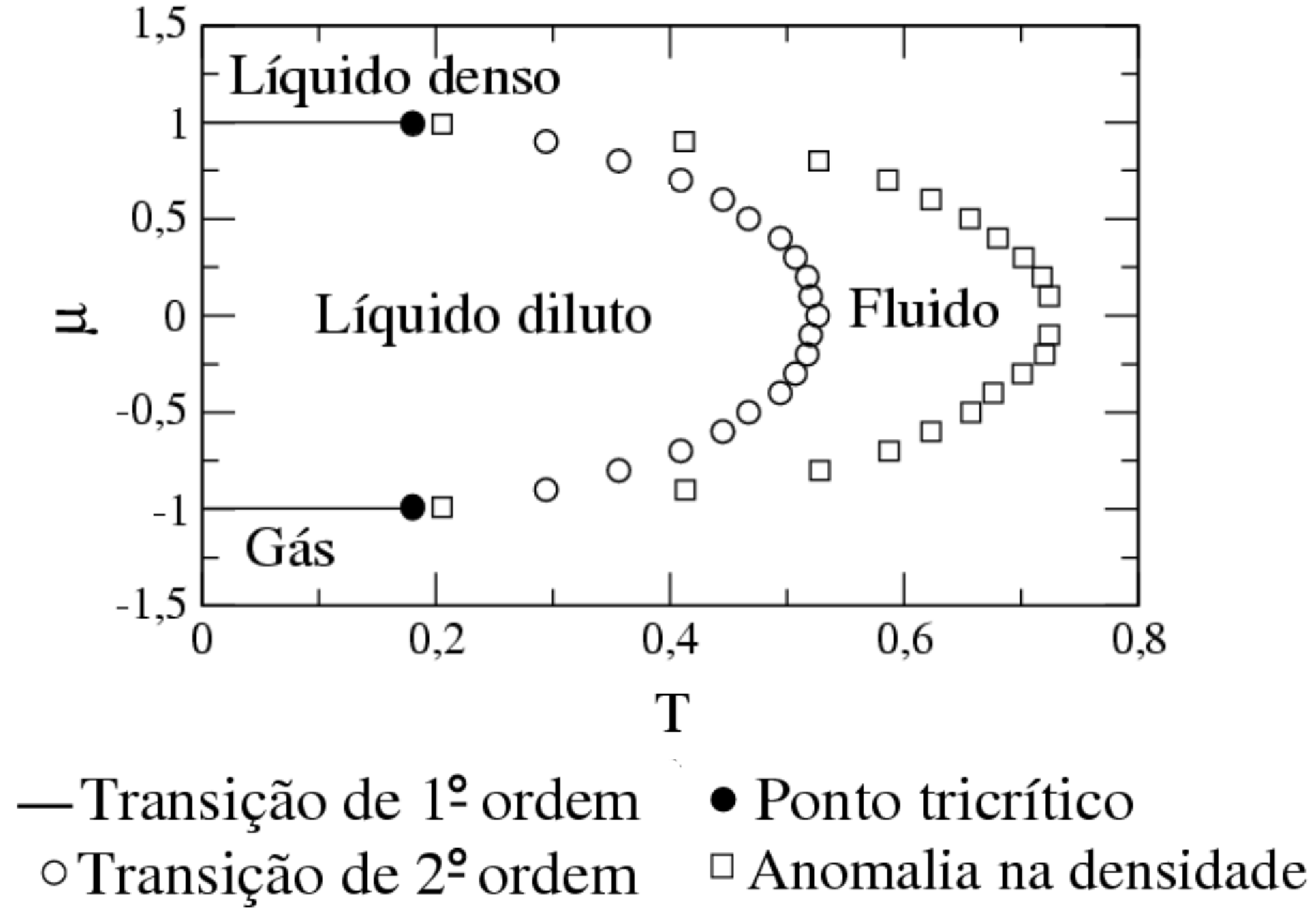}
\end{center}
\caption{$\mu^{*}\times T^{*}$ phase-diagram \label{mu.vs.T}: the 
circles indicates the critical line, the solid lines show the first-order 
transitions, the squares locate the temperature of maximum density and the 
full circles are the tricritical points. The error bars are smaller than the symbols. }
\end{figure}

Fig. \ref{d-total} shows that for  a fixed chemical potential 
the density has a maximum at a certain temperature 
$T_{max}$. The temperature of maximum density 
as a function of the chemical potential is 
 is illustrated on Fig. \ref{mu.vs.T}. 

%%%%%%%%%%%%%%%%%%%%%%%%%%%%%%%%%%%%%%%%%%%%%%%%%%%%%%%%%%%%%%%%%%%
%%%%%%%%%%%%%%%%%%%%%%%%%%%%%%%%%%%%%%%%%%%%%%%%%%%%%%%%%%%%%%%%%%

\begin{figure}
\begin{center}
\includegraphics[clip,scale=0.7]{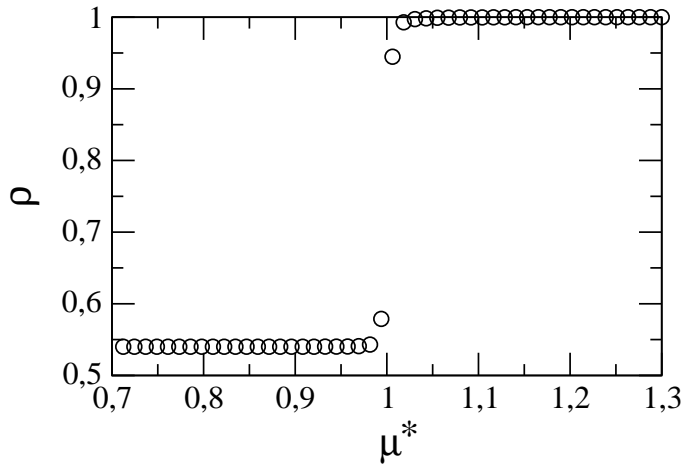}
\caption{$\rho\times\mu^*$ for $T^{*}=0.19$  \label{primeira-ordem}}
\end{center}
\end{figure}

In order to observe  the first-order phase transitions 
between the gas and SDL  phase and  between the 
SDL phase and the DL phase, simulations at fixed temperature
and  varying the 
chemical potential were performed. 
Figure \ref{primeira-ordem} shows that for 
$T^*=0.19$ the transition between the SDL and 
the DL happens at $\mu^*=1$.  At low chemical potentials
similar transition is observed between the gas and the 
SDL phases.
At the  intersection between the  critical line and the
first-order phase transition lines arise two  tricritical points
as indicated in Fig. \ref{mu.vs.T}.

%%%%%%%%%%%%%%%%%%%%%%%%%%%%%%%%%%%%%%%%%%%%%%%%%%
%%%%%%%%%%%%%%%%%%%%%%%%%%%%%%%%%%%%%%%%%%%%%%%%%%

\begin{figure}
\begin{center}
\includegraphics[clip, scale=0.4]{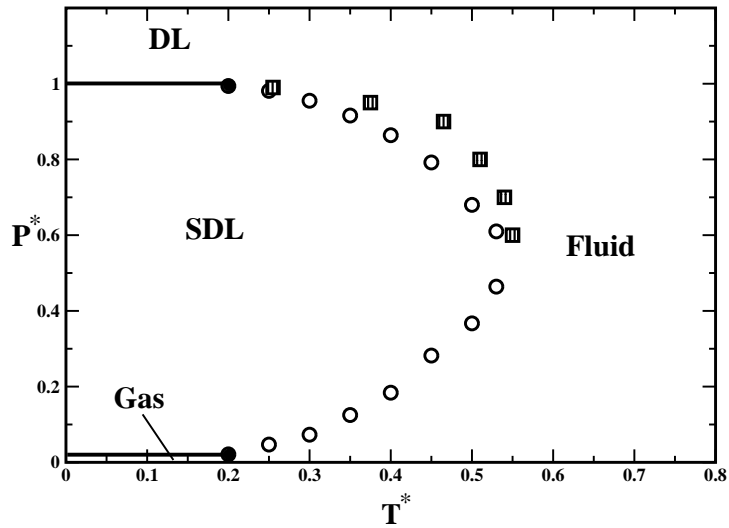}
\end{center}
\caption{$p^{*}\equiv p/|V_1|\times T^{*}$ phase diagram \label{p.vs.T}: the
solid lines indicate first-order transitions, the empty circles locate the
continuous line, the squares show the temperature of maximum density and 
the filled circles are the tricritical points.
The error bars are smaller than the symbols. }
\end{figure}

By integrating the  the Gibbs-Duhem relation:
%%%%%%%%%%%%%%%%%%%%%%%%%%%%%%%%%%%%%%%%%%%%%%%%%%%%%%%%%%%%%%%%
\begin{equation}
SdT-Vdp+Nd\mu=0,
\label{eq:gibbs-duhem}
\end{equation}
%%%%%%%%%%%%%%%%%%%%%%%%%%%%%%%%%%%%%%%%%%%%%%%%%%%%%%%%%%%%%%%%
where $S$ stands for entropy and  $p$ for pressure, 
together with the MC simulations at a fixed temperature, the dependence
of the pressure on the 
density   for
a fixed temperature was obtained. Comparing this dependence for
different temperatures, was possible to extract the phase-diagram
illustrated at Fig.\ref{p.vs.T}. Here the density anomaly appears 
as the temperature of maximum density for a fixed pressure.

%%%%%%%%%%%%%%%%%%%%%%%%%%%%%%%%%%%%%%%%%%%%%%%%%%%%%%%%%%%%%%%%%
%%%%%%%%%%%%%%%%%%%%%%%%%%%%%%%%%%%%%%%%%%%%%%%%%%%%%%%%%%%%%%%%

The phase-diagram for other values of $V_1<-2V_2$ is similar to the
one we present here. In the case of $V_1>-2V_2$ both the two
liquid phases and the density anomaly are not present.

%%%%%%%%%%%%%%%%%%%%%%%%%%%%%%%%%%%%%%
%%%%%%%%%%%%%%%%%%%%%%%%%%%%%%%%%%%%%%
\section {Conclusions}
\label{5}
%%%%%%%%%%%%%%%%%%%%%%%%%%%%%%%%%%%%%%
%%%%%%%%%%%%%%%%%%%%%%%%%%%%%%%%%%%%%%

The phase-diagram of a two dimensional lattice
gas model with competing interactions
and  in contact with a reservoir of particles and
temperature was investigated 
 with the purpose of testing this attraction/repulsion
potential for the presence  of criticality and density anomaly .

This system exhibits two liquid phases and a line
of density anomalies. Differently from the general belief, the
density anomaly line in the present case is not associated to a single critical
point \cite{Po92} but with a line
of critical points.
Besides, the density anomaly does   not arise from
a softened 
core potential but from an outer
shell repulsion that competes with a short-range attraction. We 
believe that the common ingredient that 
leads to the presence of demixing between two liquid phases
both in softened core potentials and in the present model is
the existence of two competing scales for the  interaction
 \cite{Ba04}\cite{He04}.

The relation between the form of the potential, criticality and 
the density anomaly goes as follows. The presence of two liquid
phases comes from the presence of competing scales. While 
the short-range attraction favors the formation of a dense liquid
phase, the outer shell  repulsion favors the formation of an open
structure, the SDL phase. 

In systems dominated by short-range attractive forces, if the
pressure is kept fixed,  the density increases  on cooling.
In our case, similar behavior
is only observed at high temperatures where short-range interactions
are dominant. As
the temperature is decreased,   the outer shell  repulsion  prevents 
the density to increase beyond a certain limit.
Therefore, the same competition responsible for the appearance of
two liquid phases leads to the density anomaly.

One should point out that the presence
of a  critical line instead of a single critical point as
one could generally expect \cite{Po92} is not surprising.
Due to the lattice structure, the SDL is not one single phase
but the region where two  different phases coexist ( empty/full rows and
empty/full columns). These two phases become critical together with the DL phase at the  tricritical point that is also the locus where critical line
ends.

\vspace*{1.25cm}

\noindent \textbf{\large Acknowledgments}{\large \par}

\vspace*{0.5cm} This work was supported by the Brazilian science agencies CNPq,
FINEP, Capes  and Fapergs.

\end{document}